\documentclass[aps,prc,twocolumn,showpacs,groupedaddress,floatfix]{revtex4}

\usepackage{graphicx}
\usepackage{dcolumn}

\begin{document}

\title{Hartree-Fock-Bogoliubov Calculations in Coordinate Space:\\
       Neutron-Rich Sulfur, Zirconium, Cerium, and Samarium Isotopes\\}

\author{V.E. Oberacker}
  \affiliation{Department of Physics and Astronomy, Vanderbilt University, 
      Nashville, Tennessee 37235, USA}

\author{A.S. Umar}
  \affiliation{Department of Physics and Astronomy, Vanderbilt University, 
      Nashville, Tennessee 37235, USA}

\author{E. Ter\'an}
 \altaffiliation{Physics Department, San Diego State University, San Diego, California 92182}
  \affiliation{Department of Physics and Astronomy, Vanderbilt University, 
Nashville, Tennessee 37235, USA}

\author{A. Blazkiewicz}
\affiliation{Department of Physics and Astronomy, Vanderbilt University, 
Nashville, Tennessee 37235, USA}

\date{\today}


\begin{abstract}
Using the Hartree-Fock-Bogoliubov (HFB) mean field theory in coordinate space,
we investigate ground state properties of the sulfur isotopes from the line of
stability up to the two-neutron dripline ($^{34-52}S$). In particular, we 
calculate two-neutron separation energies, quadrupole moments, and rms-radii for
protons and neutrons. Evidence for shape coexistence is found in the very
neutron-rich sulfur isotopes. We compare our calculations with results from
relativistic mean field theory and with available experimental data.
We also study the properties of neutron-rich zirconium ($^{102,104}Zr$), cerium
($^{152}Ce$), and samarium ($^{158,160}Sm$) isotopes which exhibit very large
prolate quadrupole deformations.
\end{abstract}
\maketitle


\section{\label{sec:intro}Introduction\protect\\}
One of the fundamental questions of nuclear structure physics is: how many neutrons
or protons can we add to a given nuclear isotope before it becomes unstable against
spontaneous nucleon emission? The neutron-rich side of the nuclear chart,
in particular, exhibits thousands of nuclear isotopes still to be explored 
with new Radioactive Ion Beam facilities \cite{DOE02}.
Another limit to stability is the superheavy element region around $Z=124-126$ and $N=184$
which is formed by a delicate balance between strong Coulomb repulsion and additional
binding due to closed shells \cite{KB00}. 
Theoretically, one expects profound differences between the known isotopes
near stability and exotic nuclei at the neutron dripline, e.g. 
the appearance of neutron halos and neutron skins, and large pairing correlations.

There are various theoretical approaches to the nuclear many-body problem.
For the lightest nuclei, e.g. $^{12}C$, an exact diagonalization of the Hamiltonian in
a shell model basis is feasible \cite{NV00,NO02}. Stochastic methods like
the shell model Monte Carlo approach \cite{KDL97,PD03} may be used for medium-mass 
nuclei up to $A \sim 60$. For heavier nuclei, theorists tend to
utilize self-consistent mean field theories; both non-relativistic versions
\cite{WS94,DN96,TH96,CB98,RD99,SD00,YM01,TOU03} and relativistic versions
\cite{Ri96,LR99,KB00} have been developed. 
As long as the pairing interaction is relatively weak, it is permissible to
treat the mean field and the pairing field separately via Hartree-Fock theory
with added BCS or Lipkin/Nogami pairing. This works well near the line of
stability \cite{WS94}. However, as one approaches
the driplines, pairing correlations increase dramatically
and it is essential to treat both the mean field and the pairing field
selfconsistently within the Hartree-Fock-Bogoliubov (HFB) formalism \cite{YM01}. 
While the HF(B) theories describe the ground state properties of nuclei,
their excited states can be obtained with the (quasiparticle) random phase
approximation (Q)RPA \cite{Ma01,BD02,IJ03}.

In this paper we study the ground state properties of neutron-rich even-even nuclei
up to the two-neutron dripline. Besides the large pairing correlations already
mentioned, HFB calculations face another problem in this region: 
not only does one have to consider ``well-bound'' single-particle states 
(which determine the structure near stability), but in addition there
are occupied ``weakly-bound'' states with large spatial extent. Furthermore,
because the Fermi energy for neutrons $\epsilon_F \approx 0$ at the dripline,
virtual excitations into the continuum states become important for a 
proper description of the HFB ground state. All of these features 
represent major challenges for the numerical solution.

Traditionally, the HFB equations have been solved by expanding the quasiparticle
wavefunctions in a harmonic oscillator basis \cite{ER93}. This works very well near
the line of $\beta-$ stability because only ``well-bound'' states need
to be considered. However, as one approaches the driplines, the numerical
solution becomes more challenging: in practice, it is very difficult to represent continuum
states as superpositions of bound harmonic oscillator states because the former
show oscillatory behavior at large distances while the latter decay exponentially.
On the other hand, a direct solution of the HF(B) equations on a finite-size
coordinate space lattice does not suffer from the above-mentioned shortcomings because no region
of the spatial lattice is favored over any other region: well-bound, weakly-
bound and (discretized) continuum states can be represented with the
same accuracy. Therefore, the spatial lattice representation
has inherent advantages for the theoretical description of exotic nuclei.
  
Using our recently developed HFB lattice code for deformed nuclei far from
stability \cite{TOU03}, we have investigated the ground state properties of
the sulfur isotope chain, starting at the line of stability
($N=16$) up to the two-neutron dripline (which turns out to be $N=36$ in our
HFB calculations). Our calculations show both spherical and 
quadrupole- deformed g.s. deformations; in addition, there is evidence
for shape coexistence in the very neutron-rich region. In particular, we 
calculate two-neutron separation energies, quadrupole moments, and rms-radii for
protons and neutrons. Our HFB calculations are compared with results from
relativistic mean field theory and with available experimental data.

We have also carried out HFB calculations for some recently measured heavier
systems: among medium and heavy nuclei, $^{104}$Zr ($\beta_2=0.45(4)$) and
$^{158}$Sm ($\beta_2=0.46(5)$) are among the most deformed isotopes \cite{HR03}.
The large deformation could have its
origin in the high spin down-sloping orbitals near $Z=38,40,62$ and
$N=40,64,96$. These large prolate deformations at $^{104}$Zr and $^{158}$Sm are
confirmed by Hartree-Fock-Bogoliubov calculations carried out in the
present work.


\section{\label{sec:hfb_eqns}HFB equations in coordinate space}

Recently, we have solved for the first time the HFB continuum problem
in coordinate space for deformed nuclei in two spatial dimensions without
any approximations, using Basis-Spline methods \cite{TOU03}.
The novel feature of our HFB code is
that it is capable of generating high-energy continuum states with an equivalent
single-particle energy of hundreds of MeV. In fact, early 1-D calculations
for spherical nuclei \cite{DN96} and our recent 2-D HFB calculations
have demonstrated that one needs continuum states with an equivalent
single-particle energy up to 60 MeV to describe the ground state properties
accurately near the neutron dripline. Moreover, recent QRPA calculations
by Terasaki et al. \cite{Te03} suggest that one needs to consider continuum
states up to $150$ MeV for the description of collective excited states.
It should be mentioned that current 3-D HFB codes in coordinate space,
e.g. Ref. \cite{TH96,YM01}, utilize
an expansion of the quasiparticle wavefunctions in a truncated HF-basis 
which is limited to continuum states up to about $5$ MeV of excitation energy.
Alternatively, an expansion in a {\emph stretched} oscillator basis has
also been explored \cite{SD00}.

A detailed description of our theoretical method has been published in
ref. \cite{TOU03}; in the following, we give a brief summary.
In coordinate space representation, the HFB Hamiltonian and the quasiparticle
wavefunctions depend on the distance vector ${\bf r}$, spin projection
$\sigma = \pm \frac{1}{2}$, and isospin projection $q = \pm \frac{1}{2}$
(corresponding to protons and neutrons, respectively).
In the HFB formalism, there are two types of quasiparticle wavefunctions,
$\phi_1$ and $\phi_2 $, which are bi-spinors of the form
\begin{equation}
\phi^q_{1,\alpha} ({\bf r}) =
\left(
\begin{array}{c}
\phi^q_{1,\alpha} ({\bf r,\uparrow})\\  
\phi^q_{1,\alpha} ({\bf r,\downarrow})
\end{array}
\right) , \ \ 
\phi^q_{2,\alpha} ({\bf r}) =
\left(
\begin{array}{c}
\phi^q_{2,\alpha} ({\bf r,\uparrow})\\  
\phi^q_{2,\alpha} ({\bf r,\downarrow})
\end{array}
\right) .
\end{equation}

The quasiparticle wavefunctions determine the normal density 
$\rho_q({\bf r})$ and the pairing density $\tilde \rho_q({\bf r})$
as follows
\begin{eqnarray}
\rho_q({\bf r}) \ &=& \ \sum_{E_\alpha > 0}^{\infty} \sum_{\sigma = -\frac{1}{2}}^{+\frac{1}{2}} 
       \phi^q_{2,\alpha} ({\bf r} \sigma) \ \phi^{q \ *}_{2,\alpha} ({\bf r} \sigma) \ ,
\label{eq:density}    \\
\tilde{\rho_q}({\bf r}) \ &=& \ - \sum_{E_\alpha > 0}^{\infty} \sum_{\sigma = -\frac{1}{2}}^{+\frac{1}{2}}
       \phi^q_{2,\alpha} ({\bf r} \sigma) \ \phi^{q \ *}_{1,\alpha} ({\bf r} \sigma) \ . 
\label{eq:pairing_density} 
\end{eqnarray}
In the wavefunctions, the dependence on the quasiparticle energy $E_\alpha$ is
denoted by the index $\alpha$ for simplicity.

In the present work, we use Skyrme effective N-N interactions in the p-h channel, 
and a delta interactions in the p-p channel.
For these types of effective interactions, the particle mean field Hamiltonian $h$ and the
pairing field Hamiltonian $\tilde h$ are diagonal in isospin space and
local in position space, 
\begin{equation}
h({\bf r} \sigma q, {\bf r}' \sigma' q') \ = \ \delta_{q,q'}
      \ \delta({\bf r} - {\bf r}') h^q_{\sigma, \sigma'}({\bf r}) 
\end{equation}
and
\begin{equation}
\tilde{h}({\bf r} \sigma q, {\bf r}' \sigma' q') \ = \ \delta_{q,q'}
      \ \delta({\bf r} - {\bf r}') \tilde{h}^q_{\sigma, \sigma'}({\bf r}) \ .
\end{equation}
and the HFB equations have the following structure in spin-space \cite{TOU03}:
\begin{equation}
\left( 
\begin{array}{cc}
( h^q -\lambda ) & \tilde h^q \\
\tilde h^q & - ( h^q -\lambda ) 
\end{array}
\right)
\left(
\begin{array}{c}
\phi^q_{1,\alpha} \\  
\phi^q_{2,\alpha}
\end{array}
\right)
 = E_\alpha
\left( 
\begin{array}{c}
\phi^q_{1,\alpha} \\  
\phi^q_{2,\alpha}
\end{array}
\right)
\label{eq:hfbeq2}
\end{equation}
with
\begin{equation}
h^q ({\bf r}) =
\left( 
\begin{array}{cc}
h^q_{\uparrow \uparrow}({\bf r}) & h^q_{\uparrow \downarrow}({\bf r}) \\
h^q_{\downarrow \uparrow}({\bf r}) & h^q_{\downarrow \downarrow}({\bf r})
\end{array}
\right)
, \  
\tilde h^q ({\bf r}) =
\left( 
\begin{array}{cc}
\tilde h^q_{\uparrow \uparrow}({\bf r}) & \tilde h^q_{\uparrow \downarrow}({\bf r}) \\ 
\tilde h^q_{\downarrow \uparrow}({\bf r}) & \tilde h^q_{\downarrow \downarrow}({\bf r})
\end{array}
\right) \ \ .
\label{hspin}
\end{equation}

The quasiparticle energy spectrum is discrete for $|E|<-\lambda$
and continuous for $|E|>-\lambda$ \cite{DN96}. For even-even nuclei it is customary to 
solve the HFB equations for positive quasiparticle energies and consider all negative
energy states as occupied in the HFB ground state.


\section{\label{sec:parameters} Numerical method}

Using cylindrical coordinates $(r,z,\phi)$, we introduce a 2-D grid
$(r_\alpha,z_\beta)$ with $\alpha = 1,...,N_r$ and $\beta = 1,...,N_z$.
In radial direction, the grid spans the region from $0$ to $r_{max}$.
Because we want to be able to treat octupole shapes, we do not assume left-right
symmetry in $z$-direction. Consequently, the grid extends from $-z_{max}$ to
$+z_{max}$. Typically, $z_{max} \approx r_{max}$ and $N_z$ $\approx$ $2 \cdot N_r$.

For the lattice representation, the wavefunctions and operators are represented 
in terms of Basis-Splines. B-Splines of order $M$, $B^M_i(x)$, are a set ($i=1,...,{\cal N}$) 
of piecewise continuous polynomial sections of order $M-1$; a special case are the
well-known finite elements which are B-Splines of order $M=2$. By using B-Splines of
seventh or ninth oder, we are able
to represent derivative operators very accurately on a relatively coarse grid with a
lattice spacing of about 0.8 fm resulting in a lattice Hamiltonian
matrix of relatively low dimension. While our current 2-D lattices are linear, a major
advantage of the B-Spline technique is that it can be extended to nonlinear
lattices (e.g. exponentially increasing) \cite{KO96} which will be particularly
useful for problems where one is interested in the behavior of wavefunctions at very
large distances.

The four components ($n=1,...,4$) of the HFB bi-spinor wavefunction $\psi_n (r,z)$
are expanded in terms of a product of B-Splines
\begin{equation}
\psi_n(r_\alpha,z_\beta) = \sum_{i=1}^{{\cal N}_i} \sum_{j=1}^{{\cal N}_j}
                            B^M_i(r_\alpha) B^M_j(z_\beta) c^{ij}_n \ .
\end{equation}

We construct the derivative operators contained in the Hamiltonian with the
B-Spline Galerkin method \cite{OU99} while local potentials are
represented by the collocation method \cite{WO95,KO96}. 
The numerical solution of the HFB equations results in a set of 
quasiparticle wavefunctions at the lattice points.
The corresponding quasiparticle energy spectrum contains
both bound and (discretized) continuum states. We diagonalize the HFB Hamiltonian
separately for fixed isospin projection $q$ and angular momentum projection
$\Omega$. Note that the number of quasiparticle eigenstates is determined by the
dimensionality of the lattice HFB Hamiltonian. For fixed values of
$q$ and $\Omega$, we obtain $4 \cdot N_r \cdot N_z$ eigenstates,
typically up to 1,000 MeV.

In ref.\cite{TOU03,UOT03} we have investigated the numerical convergence of several
observables as a function of lattice box size, grid spacing, and maximum angular
momentum projection $\Omega_{max}$. In the case of spherical nuclei, our calculations
have been compared with the 1-D radial HFB results of Dobaczewski et al. \cite{DN96},
and indeed there is good agreement between the two. Production runs of our
HFB code are carried out on an IBM-SP massively parallel supercomputer using
OPENMP/MPI message passing. Parallelization is possible for different angular
momentum states $\Omega$ and isospins (p/n).


\section{Numerical results and comparison with experimental data}

In this section we present numerical results of our 2D-HFB code and compare these
to experimental data and other theoretical methods. In all of our calculations
we utilized the Skyrme (SLy4) \cite{CB98} effective N-N interaction in the p-h and h-p channel, and for
the p-p and h-h channel we use a delta interaction with the same parameter set
as in ref. \cite{TOU03}: a pairing strength of $V_0=-170.0\;  MeV fm^3$, with an equivalent
s.p. energy cutoff parameter ${\cal E}_{max}=60$ MeV. All calculations reported in this
paper were carried out with B-Spline order $M = 7$ and maximum angular momentum projection
$\Omega_{max} = \frac{21}{2}$. 


\subsection{Sulfur isotope chain up to the two-neutron dripline; shape coexistence studies}

The sulfur isotopes ($Z=16$) have been investigated several years ago by Werner et al. \cite{WS94}
using self-consistent mean field models: both Skyrme-HF and relativistic mean field (RMF) model calculations
were carried out using a simple heuristic ``constant pairing gap'' approximation.
Because of the well-known deficiencies of standard pairing theory in the exotic neutron-rich
region, we have decided to re-investigate the sulfur isotope chain, starting at the line of stability
($N=16$) up to the two-neutron dripline (which turns out to be $N=36$ in our
HFB calculations). Based on the above-mentioned earlier calculations, one may
expect a wide range of ground state deformations, and in addition there has
been some evidence for shape coexistence in this region \cite{WS94}. Because
our HFB code \cite{TOU03} has been specifically designed to describe deformed
exotic nuclei, the sulfur isotopes are expected to provide a rich testing ground for
our calculations. 

Within the single-particle shell model, one would expect nuclei such as $^{44}_{16}S_{28}$
with ``magic'' neutron number $N = 28$ to be spherical. But the mean field theories predict,
in fact, deformed intrinsic shapes as a result of ``intruder'' states. Furthermore,
in some of these isotopes shape coexistence has been predicted \cite{WS94}.
All these phenomena depend strongly on the interplay between the mean field
and the pairing field with is correctly described the the HFB theory. Furthermore,
the neutron-rich $N \approx 28$ nuclei play a crucial role in astrophysics for the nucleosynthesis
of the heavy Ca-Ti-Cr isotopes \cite{S93}.

In radial ($r$) direction, our lattice extends from $0 - 12$ fm, and in symmetry axis ($z$)
direction from $-12,...,+12$ fm, with a lattice spacing of about $0.8$ fm in the central
region. Angular momentum projections $\Omega = 1/2, 3/2, ..., 21/2$ were taken into account. 

\begin{figure}[htb]
\vspace*{0.0cm}
\includegraphics[scale=0.40]{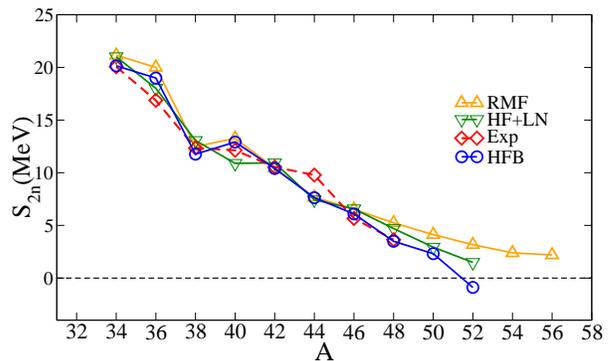}
\vspace*{0.0cm} 
\caption{\label{fig:separation_S} Two-neutron separation energies for
sulfur isotopes. The dripline is located where the separation
energy becomes zero.}
\end{figure}

Figure \ref{fig:separation_S} shows the calculated two-neutron separation energies
for the sulfur isotope chain. The two-neutron separation energy is defined as
\begin{equation}
S_{2n}(Z,N) \ = \ E_{bind} (Z,N) \ - \ E_{bind} (Z, N - 2) \; .
\end{equation}
Note that in using this equation, all binding energies must be entered with a
\emph{positive} sign. The position of the two-neutron dripline is defined by the condition
$S_{2n}(Z,N) \ = 0 $, and nuclei with negative two-neutron separation energy are
unstable against the emission of two neutrons.

The two-neutron separation energies have been calculated using various methods:
in addition to HFB calculations (i.e. selfconsistent mean field with pairing),
we have also carried out Hartree-Fock calculations with added Lipkin/Nogami
pairing (HF+LN), and we compare our results to the relativistic mean field
with BCS pairing (RMF) calculations by Lalazissis et al. \cite{LR99}.
Experimental data based on measured binding
energies \cite{AW95} are available up to the isotope $^{48}S$. Fig. \ref{fig:separation_S}
shows that both the HFB and RMF calculations are in good agreement with experiment
where available but there are dramatic differences as we approach the two-neutron
dripline: Our HFB calculations predict $^{50}S$ to be the last isotope that is
stable against the emission of two neutrons. By contrast, the RMF approach predicts
 $S_{2n}(Z,N) > 0 $ at least up to $^{56}S$. Our HF+LN calculations also yield positive
$2n$-separation energies in the mass region investigated here. It should be stressed that,
on theoretical grounds, neither BCS-type nor Lipkin-Nogami type pairing is
justified for the very neutron-rich isotopes. Furthermore, it is well-known that
the HF+LN method breaks down at magic numbers (the pairing gap does not vanish).

\begin{figure}[htb]
\vspace*{0.0cm}
\includegraphics[scale=0.40]{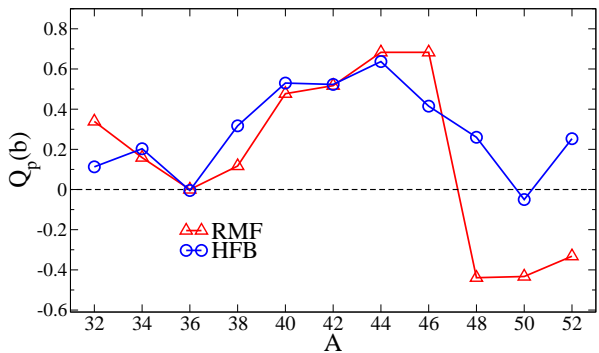}
\vspace*{0.0cm}
\caption{\label{fig:quad_p} Quadrupole moment for protons (in units of barn)
         for even-even sulfur isotopes.}
\end{figure}

In Fig. \ref{fig:quad_p} we show a comparison of the HFB and RMF results for
the intrinsic electric quadrupole moments of the sulfur isotopes. In most cases, the HFB and
RMF calculations show a similar trend: we observe a region with predominantly
prolate deformation. Note, however, that for the most neutron-rich sulfur isotopes, our HFB
theory predicts a prolate ground state whereas RMF theory yields an oblate
shape. Direct measurements of electric quadrupole moments are only available for two of
the sulfur isotopes. The data compilation of Stone \cite{S01} yields intrinsic electric
quadrupole moments of $Q^{exp}=-0.15 b$ for $^{32}S$ and $Q^{exp}=+0.04 b$ for $^{34}S$.
Our HFB code yields $Q_{HFB}= +0.113 b$ and $+0.203 b$ for these two isotopes. The RMF calculations
of Lalazissis et al. \cite{LR99} give values of $Q_{RMF}=+0.339 b$
and $+0.159 b$, respectively.

In intermediate-energy Coulomb excitation experiments, the energies
and $B(E2)$ values of the lowest excited $2^+$ state were measured
for $^{38,40,42}S$ \cite{Sch96} and for $^{44}S$ \cite{Gl97}. The analysis of the measured
$BE(2)$ values in terms of the simple quadrupole-deformed rotor model yields 
the following experimental quadrupole deformations for the even sulfur isotopes $^{38-44}S$:
$|\beta_2^{exp}| = 0.246(16),\ 0.284(16),\ 0.300(24),\ 0.258(36)$; our corresponding 
Skyrme-HFB theory results are $\beta_2^{HFB} = 0.16,\ 0.26,\ 0.25,\ 0.29$. Apparently, the
HFB results for $^{40,42,44}S$ are in good agreement with experiment,
but there is a discrepancy in the case of $^{38}S$: Skyrme-HFB predicts a
less deformed shape than the experimental value. There is an even larger
discrepancy between experiment and the RMF calculations which yield an almost
spherical shape, $\beta_2^{RMF} = 0.054$.

Both HFB and RMF calculations reveal shape coexistence in this region,
with an energy difference between the ground state and the shape isomer that is
usually quite small. For example, in the case of $^{48}_{16}S_{32}$
our HFB code yields a ground state binding energy of $-362.56$ MeV with
a quadrupole deformation of 
$\beta_2 = 0.11$, and an oblate minimum at $\beta_2 = -0.15$ which is
only $0.49$ MeV higher than the ground state. The RMF predicts in this case 
a ground state binding energy of $-362.97$ MeV with oblate deformation of
$\beta_2 = -0.25$, and a shape isomer with $\beta_2 = +0.179$ which is
located $0.29$ MeV above the ground state.

\begin{figure}[htb]
\vspace*{0.0cm}
\includegraphics[scale=0.40]{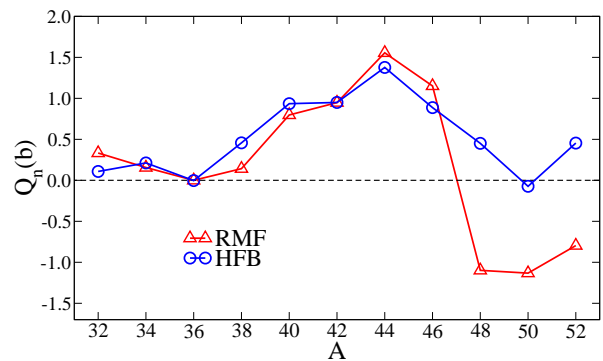}
\vspace*{0.0cm}
\caption{\label{fig:quad_n} Quadrupole moment for neutrons (in units of barn)
         for even-even sulfur isotopes.}
\end{figure}

In Fig. \ref{fig:quad_n} we show a comparison of the quadrupole moment for neutrons
predicted by our HFB calculations and the RMF results of ref.\cite{LR99}.
In both cases, the general trend is very similar to the result obtained for protons.

\begin{figure}[htb]
\vspace*{0.0cm}
\includegraphics[scale=0.40]{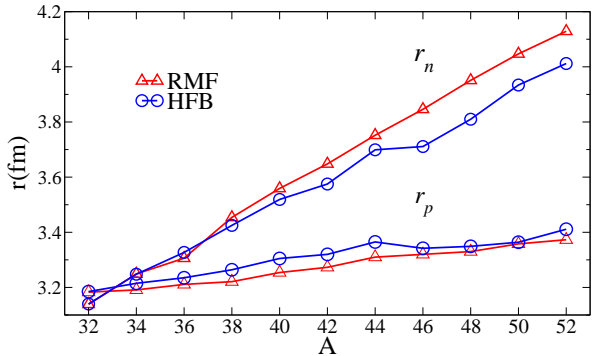}
\vspace*{0.0cm}
\caption{\label{fig:rmsrad} Root mean square radii of protons and neutrons
         for even-even sulfur isotopes.}
\end{figure}

In Fig. \ref{fig:rmsrad} we compare the root-mean-square radii of
protons and neutrons predicted by our HFB calculations and the RMF calculations
with BCS pairing \cite{LR99}. Near the line of $\beta$-stability, the proton and neutron
radii are almost identical, but as we approach the $2n$-dripline, we see clearly
the development of a ``neutron skin'' as evidenced by the large difference between the
neutron and proton rms radii. For example, in $^{50}_{16}S_{34}$ our HFB calculations
yield $r_n=3.935$ fm and $r_p=3.364$ fm, respectively. In general, the RMF calculations
predict larger neutron rms radii for mass numbers $A \geq 38$ than do our HFB calculations.

\subsection{Strongly deformed neutron-rich zirconium, cerium, neodymium, and samarium isotopes}

Recently, triple-gamma coincidence experiments have been carried out with Gammasphere
at LBNL \cite{HR03} which have determined half-lives and quadrupole deformations of
several neutron-rich zirconium, cerium, and samarium isotopes. Furthermore, laser
spectroscopy measurements \cite{Ca02} for zirconium isotopes have yielded
precise rms-radii in this region. These medium/heavy mass nuclei are among the most
neutron-rich isotopes ($N/Z \approx 1.6$) for which spectroscopic data are available.
It is therefore of great interest to compare these data with the predictions of the
selfconsistent HFB mean field theory. 

A comparison of our HFB results and experimental data is given in table \ref{table:zrcesm}.
The theoretical quadrupole deformations of the proton charge distributions agree
very well with the measured data of ref.\cite{HR03}. In addition, our calculated
proton rms-radius for $^{102}Zr$ is in good agreement with recent laser spectroscopic
measurements (see Fig.4 of ref.\cite{Ca02}). Theoretical HFB predictions are also given
for the neutron density distributions. 

\begin{table}[hbt!] 
\caption{\label{table:zrcesm}
Our HFB results for neutron-rich zirconium, cerium, neodymium, and samarium isotopes.
The first column lists the neutron-to-proton ratio $N/Z$. Subsequent columns 
display quadrupole deformations $\beta_2(p),\beta_2(n)$ and rms-radii
$r_p,r_n$ of protons and neutrons. Recent experimental data for quadrupole
deformations are taken from ref.\cite{HR03}. The rms-radius for $^{102}Zr$ 
was measured in ref.\cite{Ca02}.}

\begin{ruledtabular}
\begin{tabular}{ l c c c c c c c}
             & $N/Z$  & $\beta_2(p)$  & $\beta_2^{exp}(p)$  & $\beta_2(n)$ & $r_p$ (fm) & $r_p^{exp}$ (fm) & $r_n$ (fm)\\
\hline 
$^{102}Zr$   & 1.55   & 0.43          & 0.42(5)             & 0.43         & 4.47       & 4.54             & 4.65 \\
$^{104}Zr$   & 1.60   & 0.45          & 0.45(4)             & 0.45         & 4.49       &                  & 4.70 \\
$^{152}Ce$   & 1.62   & 0.32          & 0.30(3)             & 0.33         & 5.01       &                  & 5.22 \\
$^{156}Nd$   & 1.60   & 0.37          &                     & 0.36         & 5.08       &                  & 5.27 \\
$^{160}Sm$   & 1.58   & 0.38          &                     & 0.37         & 5.13       &                  & 5.31 \\
\end{tabular}
\end{ruledtabular}
\end{table}

In table \ref{table:comparison_sm158} we give a more detailed comparison of
various theoretical calculations for $^{158}_{\ 62}Sm$.
As stated earlier, only the HFB theory provides a self-consistent treatment
of both mean field and pairing properties. In contrast, the HF + Lipkin/Nogami and
the RMF calculations treat mean field and pairing as separate entities. By comparing
the HFB and HF+LN results in the first two columns with the measured binding
energies and quadrupole deformations (last column) we find that our HFB lattice calculation yields
values which are closer to the experimental data. Table \ref{table:comparison_sm158}
also shows that while the RMF theory reproduces the experimental binding energy
quite well it seriously underpredicts the strong quadrupole deformation measured
in ref. \cite{HR03}. In addition, table \ref{table:comparison_sm158} compares theoretical
results for other observables  such as rms-radii for neutrons and protons, Fermi energies  
($\lambda_n, \lambda_p$), pairing gaps ($\Delta_n, \Delta_p$), and pairing energies
$E_{pair}(n), E_{pair}(p)$.

\begin{table}[hbt!] 
\caption{\label{table:comparison_sm158}
Ground state properties for $^{158}_{\ 62}Sm$. The first two columns 
give the results of our present work in the Hartree-Fock Bogoliubov theory
(HFB) and in the Hartree-Fock plus Lipkin-Nogami pairing theory (HF+LN).
The third column shows RMF theory results \cite{LR99}, and the last column
gives a comparison with recent experimental data \cite{HR03}.}
\begin{ruledtabular}
\begin{tabular}{ l c c c c}
                    & HFB     & HF+LN   & RMF      & exp  \\
\hline 
B. E.   (MeV)       &-1,290.2 &-1,286.6 &-1,291.98 & -1,291.9  \\
$\beta_2(p)$        & 0.375   &   0.359 & 0.292    &  0.46(5) \\
$r_n$ (fm)          &  5.27   &         & 5.376    &     \\
$r_p$ (fm)          &  5.15   &         & 5.098    &     \\
$\lambda_n$ (MeV)   &  -5.63  & -5.37   &          &     \\ 
$\lambda_p$ (MeV)   &  -9.21  & -8.93   &          &     \\ 
$\Delta_n$ (MeV)    & 0.31    & 0.65    &          &     \\ 
$\Delta_p$ (MeV)    & 0.37    & 0.75    &          &     \\ 
$E_{pair}(n)$ (MeV) &  -0.96  & -4.90   &          &     \\
$E_{pair}(p)$ (MeV) &  -1.19  & -4.57   &          &     \\
\end{tabular}
\end{ruledtabular}
\end{table}

\begin{figure}[!htb]
\begin{center}
\vspace*{0.5in}
\scalebox{0.5}{\includegraphics*[117pt,218pt][320pt,564pt]{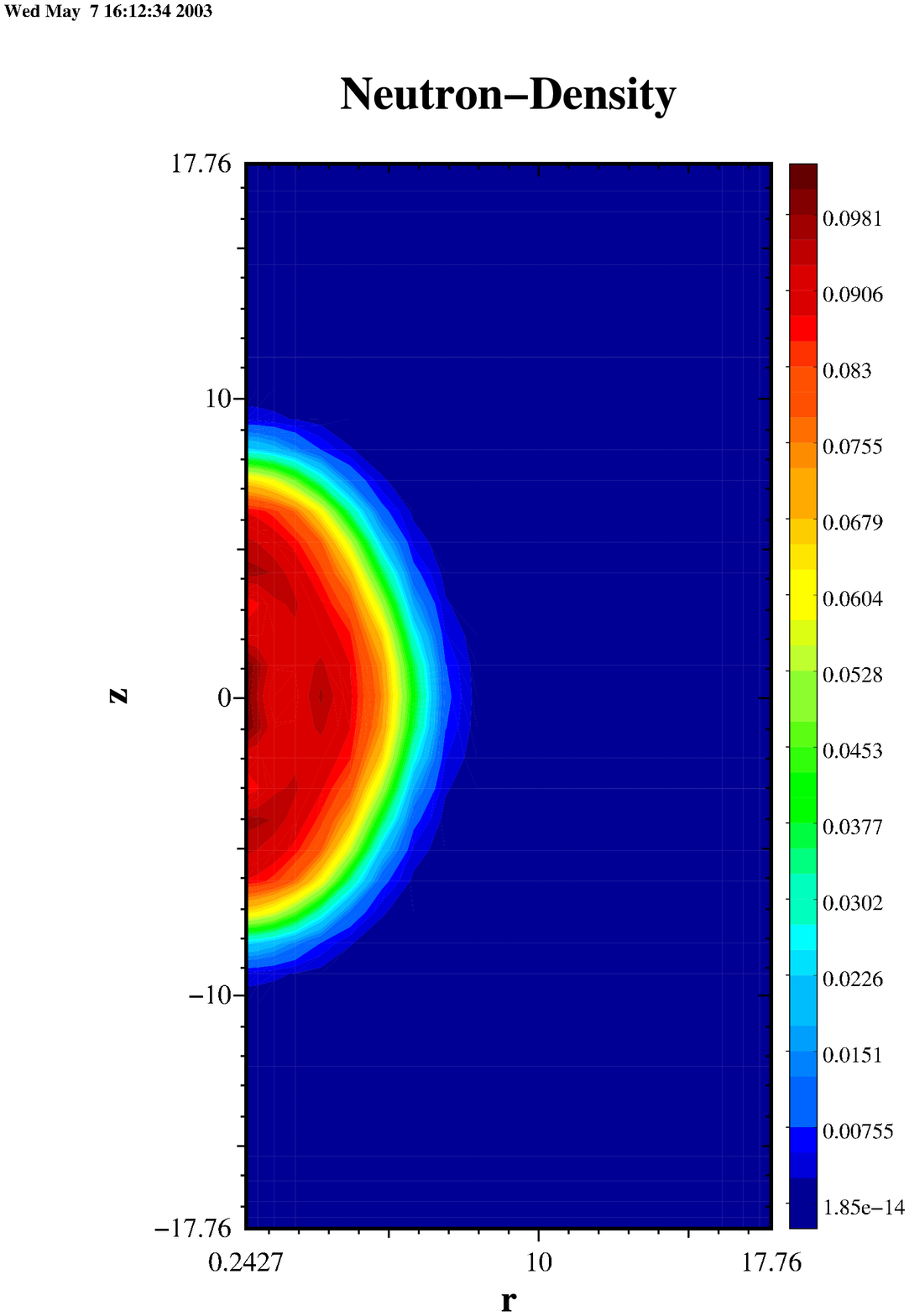} }
\hspace*{0.0in}
\scalebox{0.5}{\includegraphics*[117pt,218pt][320pt,564pt]{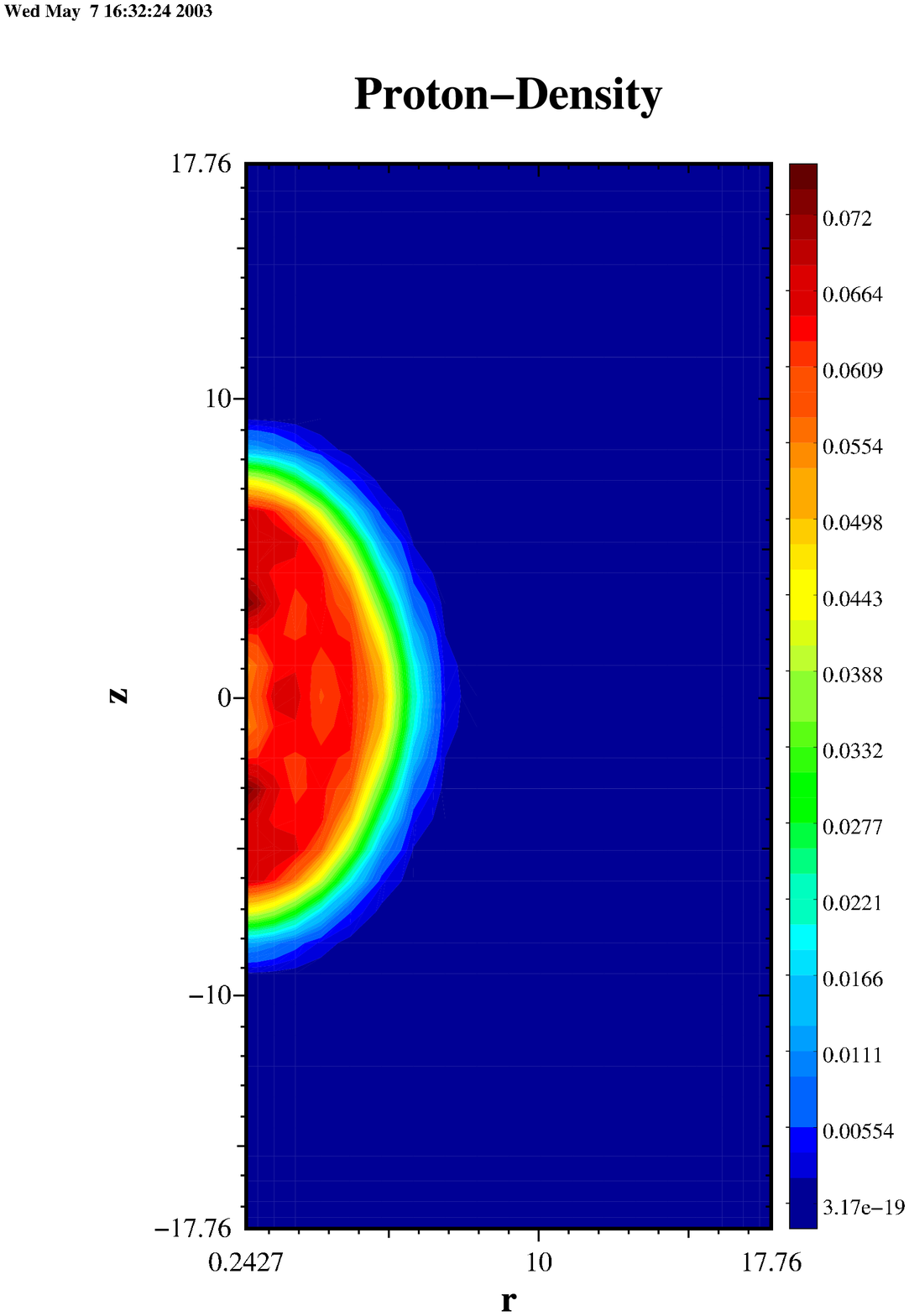} }
\vspace*{0.0in}
\caption{Density distribution for neutrons (left) and protons (right) in $^{158}_{\ 62}$Sm}
\label{sm158_dens}
\end{center}
\end{figure}

In Fig.~\ref{sm158_dens} we depict contour plots of the density distributions for
neutrons and protons in $^{158}_{\ 62}Sm$. The large prolate quadrupole deformation is clearly
visible. We also observe small density enhancements near the center of the nucleus
which are caused by the nuclear shell structure.

\begin{figure}[!htb]
\begin{center}
\vspace*{0.5in}
\scalebox{0.5}{\includegraphics*[117pt,218pt][320pt,564pt]{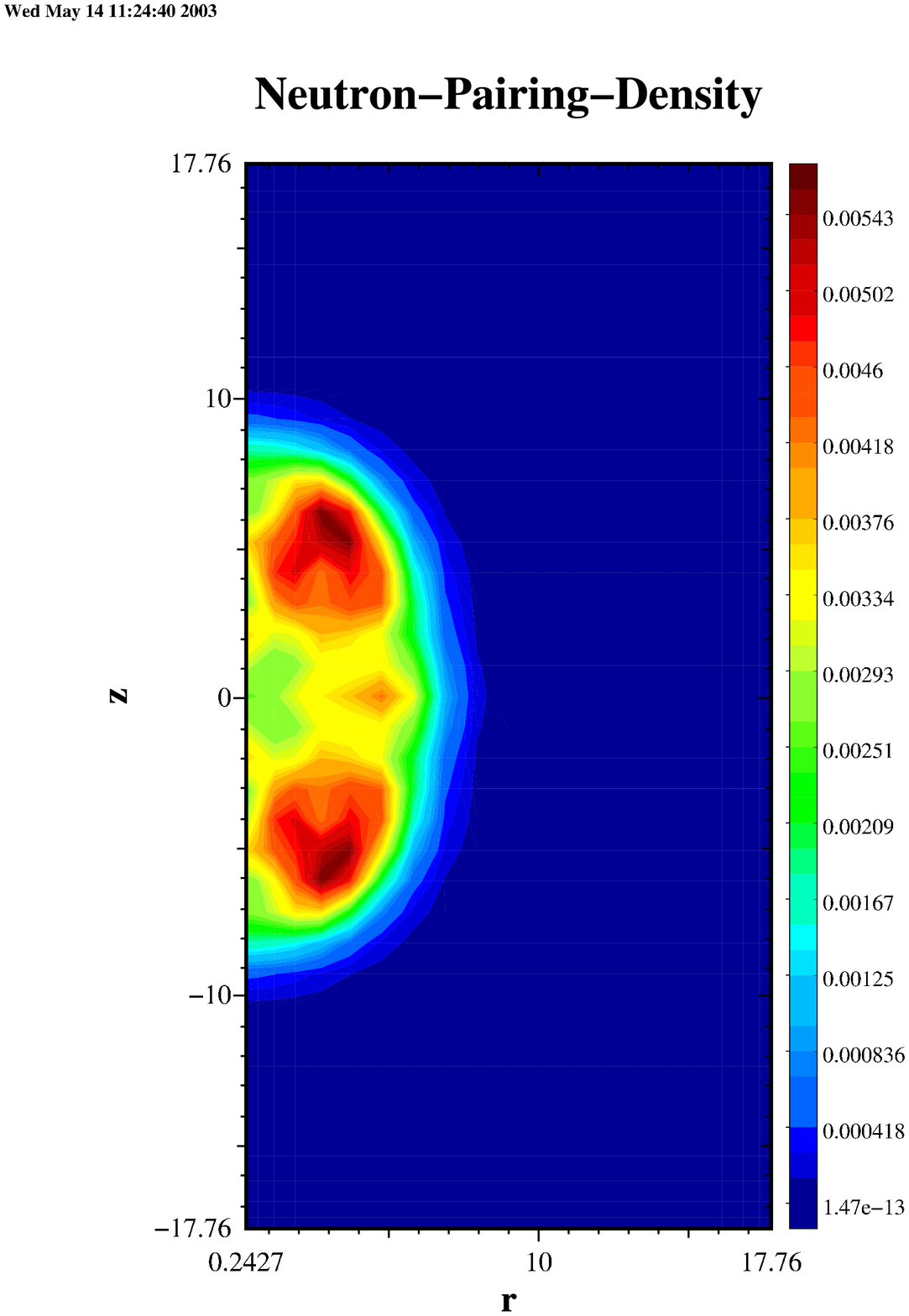} }
\hspace*{0.0in}
\scalebox{0.5}{\includegraphics*[117pt,218pt][320pt,564pt]{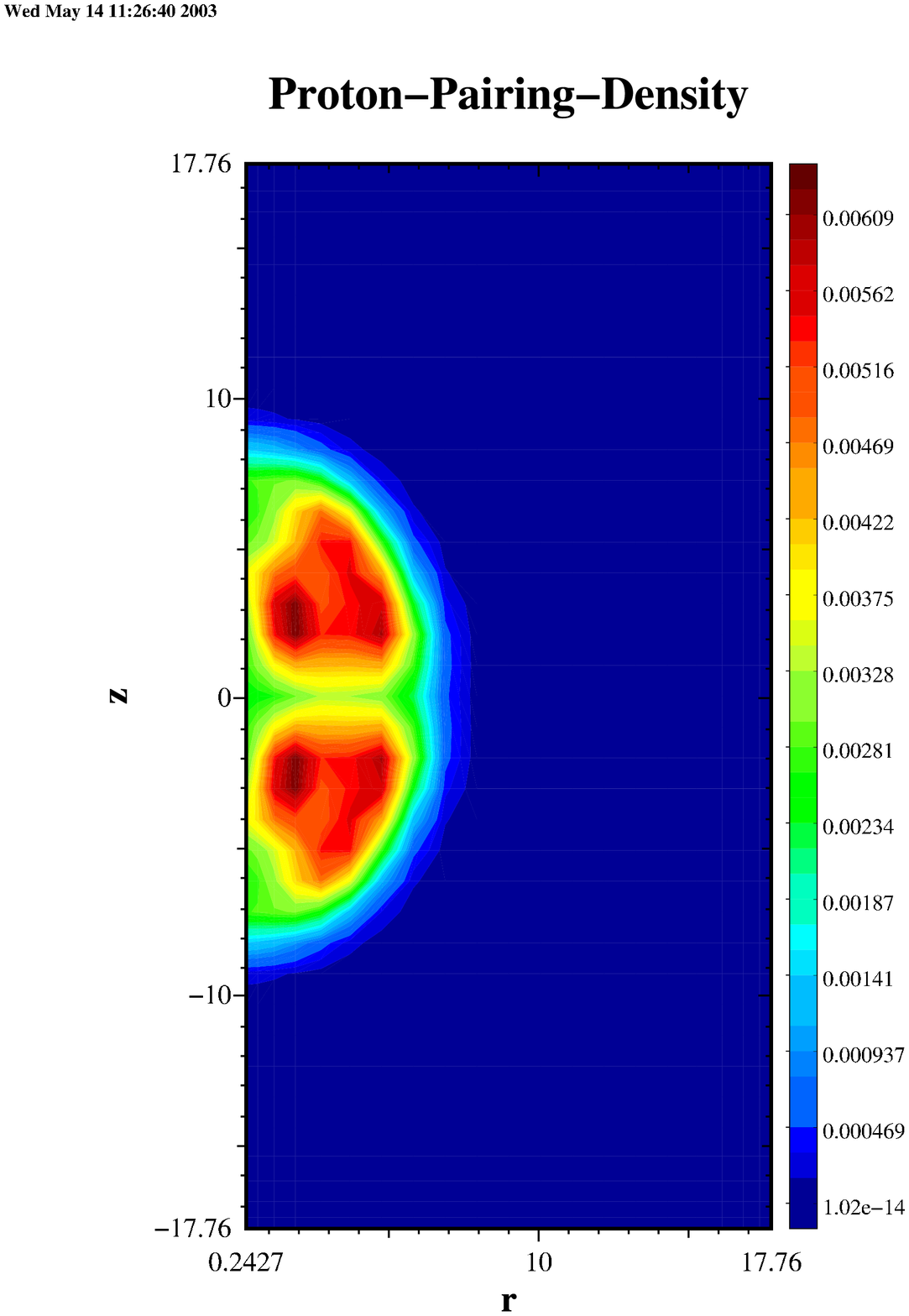} }
\vspace*{0.0in}
\caption{Pairing density distribution of neutrons (left) and protons (right) in $^{158}_{\ 62}$Sm}
\label{sm158_pairdens}
\end{center}
\end{figure}

Fig.~\ref{sm158_pairdens} shows the corresponding pairing density for
neutrons and protons; as discussed in ref.\cite{DN96}, this quantity describes the 
probability of \emph{correlated} nucleon pair formation with opposite
spin projection, and it determines the pair transfer formfactor.
We can see that most correlated pair formation in $^{158}_{\ 62}$Sm
takes place outside the central region of the nucleus.

\section{\label{sec:conclusions}Conclusions}

In this paper, we have performed Skyrme-HFB calculations in coordinate space
for several neutron-rich exotic nuclei. The coordinate space method has
the advantage that well-bound, weakly bound and (discretized) continuum
states can be represented with the same numerical accuracy. The novel feature
of our lattice HFB code is that it takes into account high-energy continuum
states with an equivalent single-particle energy of 60 MeV or more. This
feature is crucial when one studies nuclei near the neutron dripline \cite{TOU03}.

We have calculated the ground state properties of the sulfur isotope chain ($Z=16$),
starting at the line of stability ($N/Z=1$) up to the two-neutron dripline
with a neutron-to-proton ratio of $N/Z \approx 2.2$. 
In particular, we have calculated two-neutron separation energies, quadrupole
moments and rms radii for protons and neutrons. In comparing our HFB calculations
with other theoretical methods (RMF with BCS pairing and HF+Lipkin/Nogami)
we find similar results near stability but dramatic differences near the
2n-dripline (see Figures 1 - 4). For example, our HFB calculations predict
$^{50}S$ to be the last isotope that is stable against the emission of two neutrons
whereas the RMF approach predicts $S_{2n}(Z,N) > 0 $ at least up to $^{56}S$.
For $^{48}S$, the last even-even isotope for which experimental binding 
energies are available (Fig. 1), the experimental value for the two-neutron
separation energy is $3.64$ MeV, as compared to our HFB calculation result
of $3.49$ MeV and the RMF result of $5.24$ MeV.
Both our HFB calculations and the RMF calculations of ref.\cite{LR99} predict
the existence of shape isomeric states in the neutron-rich sulfur isotopes.
In Fig. \ref{fig:quad_p} we compare calculated electric quadrupole moments.
In most cases, the HFB and RMF calculations yield prolate quadrupole deformations.
However, for the most neutron-rich sulfur isotopes, our HFB
theory predicts a prolate ground state whereas RMF theory yields an oblate
shape. Shape coexistence is found both in the HFB and RMF calculations, with
fairly small energy difference between the ground state and the shape isomer.
Specific results are given for $^{48}_{16}S_{32}$. A comparison between 
the root-mean-square radii of protons and neutrons clearly exhibits the 
development of a ``neutron skin'' in the neutron rich sulfur isotopes: for
example, in $^{50}_{16}S_{34}$ our HFB calculations yield neutron and proton
rms radii of $r_n=3.935$ fm and $r_p=3.364$ fm, respectively.

In connection with recent experiments at Gammasphere, we have carried out HFB
calculations of medium/heavy mass nuclei with $N/Z \approx 1.6$. In particular,
we have examined the isotopes $^{102,104}Zr$, $^{152}Ce$, $^{156}Nd$, and
$^{158,160}Sm$. The theoretical quadrupole charge deformations for
zirconium and cerium are in very good agreement with the new data. Also, our
calculated proton rms radius for $^{102}Zr$ agrees with recent laser spectroscopic
measurements. Table \ref{table:zrcesm} gives a summary of these results and
presents predictions for two neutron-rich isotopes, $^{156}Nd$ and $^{160}Sm$,
for which experimental data are expected to become available in the near future. 


\begin{acknowledgments}
This work has been supported by the U.S. Department of Energy under grant No.
DE-FG02-96ER40963 with Vanderbilt University. The numerical calculations
were carried out at the IBM-RS/6000 SP supercomputer of the National Energy Research
Scientific Computing Center which is supported by the Office of Science of the
U.S. Department of Energy.
\end{acknowledgments}




\end{document}